\documentclass[letter]{aa}

\usepackage[utf8]{inputenc}
\usepackage{graphicx}
\usepackage[export]{adjustbox}
\usepackage{stackengine}
\usepackage{epstopdf}
\usepackage{url}
\usepackage{natbib}
\usepackage{xcolor}
\usepackage{multirow}
\usepackage{gensymb}
\usepackage{subcaption}
\usepackage{float}
\usepackage[percent]{overpic}
\usepackage{array}
\usepackage{booktabs}
\usepackage[switch]{lineno}

\newcommand\CHECK[1]{\textcolor{red}}
\newcolumntype{P}[1]{>{\centering\arraybackslash}p{#1}}

\newcommand{\PSR}{PSR\,J1135--6055}
\newcommand{\psr}{PSR\,J1135--6055~}

\newcommand{\degrees}{\ensuremath{^\circ}}
\newcommand{\hours}{\ensuremath{^\mathrm{h}}}
\newcommand{\minutes}{\ensuremath{^\mathrm{m}}}
\newcommand{\seconds}{\ensuremath{^\mathrm{s}}}

\begin{document}

    \title{Jet-like structures from PSRJ1135--6055}
    \titlerunning {Jets from PSRJ1135--6055}
    \author{ P. Bordas\inst{1} \and X. Zhang\inst{1} }
    \institute{DDepartament de Física Quàntica i Astrofísica, Institut de Ciències del Cosmos (ICCUB), Universitat de Barcelona (IEEC-UB), Martí i Franquès 1, 08028 Barcelona, Spain. }
    \date{Received --- ; accepted ---}


\abstract{Pulsar wind nebulae (PWNe) produced from supersonic runaway pulsars can render extended X-ray structures in the form of tails and prominent jets. In this Letter we report on the analysis of $\sim$130~ks observations of the PWN around \psr obtained with the \textit{Chandra} satellite. The system displays bipolar jet-like structures of uncertain origin, a compact nebula around the pulsar likely formed by the bow shock ahead of it, and a trailing tail produced by the pulsar fast proper motion. The spectral and morphological properties of these structures reveal strong similarities with the PWNe in other runaway pulsars like PSR J1509--5850 and Geminga. We discuss their physical origin considering both canonical PWN and jet formation models as well as alternative scenarios that can also yield extended jet-like features following the escape of high-energy particles into the ambient magnetic field.}


\maketitle

\section{Introduction}

The interaction of pulsar wind nebulae (PWNe) with the surrounding interstellar medium (ISM) can render distinct morphological features, including torus-like structures and bipolar jets in the case of pulsars displaying slow proper motions, and bow-shaped shocks and extended cometary tails when the pulsar is moving at high velocities (see e.g. \citet{Gaensler2006}). In the X-ray band, these structures have been resolved in detail in a number of cases (\citealp{Kargaltsev2008}). A small subset of supersonically moving PWN (sPWN from now on) have in addition been observed to display unusually long X-ray outflows. The most prominent examples of such outflows are found in the Guitar Nebula \citep{Hui2012} and in the Lighthouse Nebula \citep{Pavan2014}, but the number of cases is gradually increasing (\citealp{Kargaltsev2017}). The origin of these extended jet-like structures is unknown, but their length, almost-rectilinear geometry, and misaligned orientation with respect to the pulsar proper motion are difficult to be accommodated in the framework of canonical pulsar jet formation theories.
This led to alternative interpretations, including a scenario in which particles accelerated at the PWN/medium shock interface escape the PWN and diffuse into the ambient medium magnetic field, yielding synchrotron emission which is then observed as quasi-rectilinear jet-like structures (see \citealp{Bandiera2008}). Recent numerical simulations seem to support such scenario, explaining also the asymmetric morphologies observed at both sides of the pulsar \citep{Barkov2019a}. Deep observations of known sources and the discovery of similar jet-like structures from new systems are needed to understand their origin. \psr is a new sPWN candidate displaying extended jet-like features.

\psr was discovered with the \textit{Fermi}-LAT in a blind search as an energetic gamma-ray only pulsar \citep{Deluca2011}.
It is a young ($\tau \sim 10^{4}$~yr) and energetic pulsar ($\dot{E} \gtrsim 10^{36}$~erg~s$^{-1}$) located at $d \sim 2.8$~kpc and displaying a high proper motion velocity $v_{\perp} \lesssim 330$~km~s$^{-1}$ (\citealp{Kargaltsev2017}). The analysis of archival \textit{Chandra} data obtained from the observation of the supernova remnant (SNR) G293.8+0.6, covering also \PSR, revealed for the first time the presence of both the pulsar and its nebula in the X-ray band \citep{Marelli2012}. The SNR G293.8+0.6, likely associated with \PSR, was on the contrary not detected. The morphological properties of the extended emission around \psr obtained with these $\sim 37$~ks \textit{Chandra} observations unveiled two prominent large-scale jet-like structures at both sides of the pulsar.


In this Letter we report on the analysis of archival \textit{Chandra} observations of \PSR, amounting to a total of about $\sim 130$~ks exposure time. We focus on the morphological and spectral properties of this sPWN to constrain the nature of its jet-like outflows. We compare our results with those reported for other sPWNe, finding strong similarities with the PWNe around PSR J1509--5850 and Geminga. Our results are interpreted in the general framework of pulsar jet formation theory and in alternative scenarios accounting for the production of jet-like features from sPWN.

\section{\textit{Chandra} observations, data analysis and results}
\label{observations}

\noindent \textit{Chandra} observed the field-of-view covering \psr in three occasions. The first data set (DS-1) was aimed at the study of the nearby SNR G293.8+0.6. These observations revealed the presence of \psr and its PWN in X-rays for the first time. Further observations with \textit{Chandra} were focused on the PWN itself, for a total of $\sim 90$\,ks for DS-2 and DS-3). A summary of the \textit{Chandra} observations of \psr  is given in Table~\ref{table:observations}.
%
%
\begin{table}[t]
\begin{center}
\begin{tabular}{cccc}
\hline
Data set & Obs. ID & Date & Exposure(ks)  \\
\hline
1 & 3924 & 2003-08-24 & 36.11  \\
2 & 15966 & 2014-12-30 & 55.54 \\
3 & 17572 & 2015-01-03 & 35.73 \\
\hline
\end{tabular}
\end{center}
\caption{\textit{Chandra} observations covering \psr used for the analysis reported here.
data set 1 (DS-1) aimed for the study of SNR\,G293.8+0.6. DS-2 and DS-3 correspond to subsequent dedicated \textit{Chandra} observations of \psr. Columns indicate archival Observation ID, data taking dates, and total observation exposure time.}
\label{table:observations}
\end{table}
%
%
\noindent The three data sets correspond to observations performed in \textit{Chandra}'s \texttt{VERY FAINT} timed exposure mode using the instrument's ACIS-S detector.
Data processing was performed using the \textit{Chandra} Interactive Analysis of Observations (CIAO) software \citep{Ciao2006}, version 4.11, together \textit{Chandra}'s Calibration Database (CALDB), version 4.8.2. All observations were reprocessed using \texttt{chandra$\_$repro} in \texttt{VFAINT} background cleaning mode. Unless otherwise noted, we consider only events on S3 (ccd$\_$id=7) chip and restrict the energy range to 0.5 -- 7.0 keV.

We used the \texttt{fluximage} CIAO tool to generate broad-band exposure-corrected flux maps for each of the three data sets. We then employed the Voronoi Tessellation and Percolation source detection tool \texttt{vtpdetect} on the reprocessed level=2 event file using the previously generated exposure maps, as this method is optimized for the detection low surface-brightness extended sources \citep{Vtpdetect1993}. The generated source lists were used to remove sources in the field of view (FoV). We also checked the presence of X-ray flares by extracting and filtering the background light-curve (in temporal bins of 200~s width) for deviations larger than 3 sigma.
The effective final exposures obtained for each observation are 32.36 ks, 55.14 ks, and 35.73ks for DS-1, DS-2 and DS-3, respectively, summing up to a total of 127.38 ks (see Table~\ref{table:observations}).

We merged the three data sets to perform a morphological analysis of \psr and its surroundings. We employed the WCS reference system of DS-2 since it is the longest of the two dedicated pointings on the source and we assume that between these two pointings there is no significant difference in the astrometry. To merge DS-1 we applied the \texttt{wavdetect} tool to detect point sources in both DS-1 and DS-2. The two output source lists were then cross-matched to re-project the aspect solution file of DS-1 using \texttt{reproject$\_$aspect}. Finally, by running the CIAO tools \texttt{wcs$\_$match} and \texttt{wcs$\_$update} we ensured a common WCS reference system for the three data-sets.

Merging was performed using the CIAO routine \texttt{merge$\_$obs}, with binsize set to 1, in the broad-band energy range (0.5--7.0 keV). A map of the retrieved raw counts was then divided by the corresponding exposure map, and then normalised with the exposure time to produce the final flux map (see Fig.~\ref{skymap}).

\subsection{Results: morphology}
\label{morphology}

\begin{figure}[t]
\centering
\includegraphics[width=0.5\textwidth]{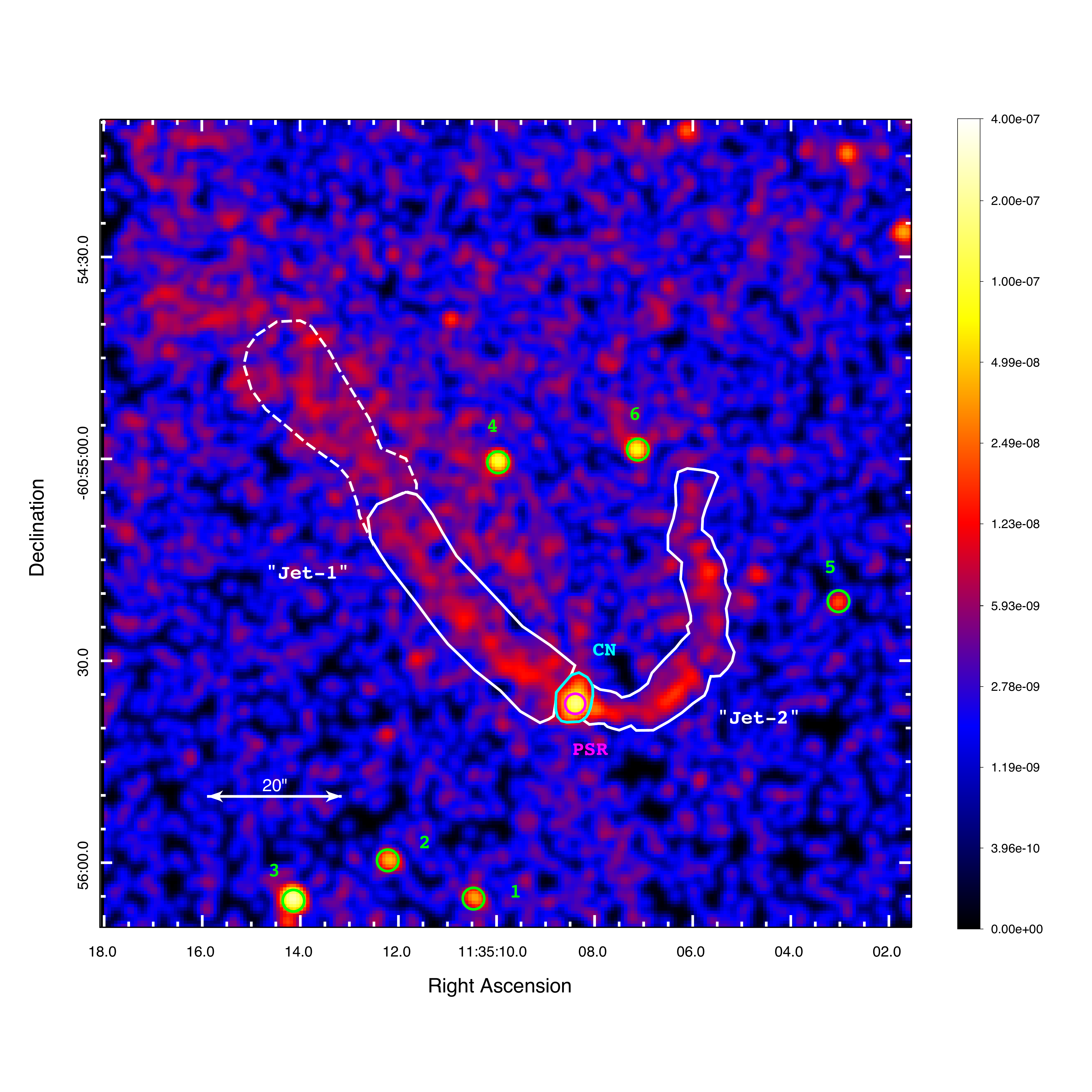}
\caption{Exposure-corrected flux map of the extended emission around \psr for energies in the range 0.5-7.0 keV. The image has been smoothed with a 1.5'' Gaussian kernel. The following regions are shown: Jet-1 inner (white eastern polygon) and outer regions (white dashed eastern polygon), Jet-2 (white western polygon), compact nebula (cyan polygon), and the pulsar J1135--6055 (magenta circle with radius of 1.5''). Also six point sources in the field of view are marked with green circles. }
\label{skymap}
\end{figure}
\begin{figure*}
\begin{subfigure}{.33\textwidth}
  \centering
  \includegraphics[width=1.0\textwidth]{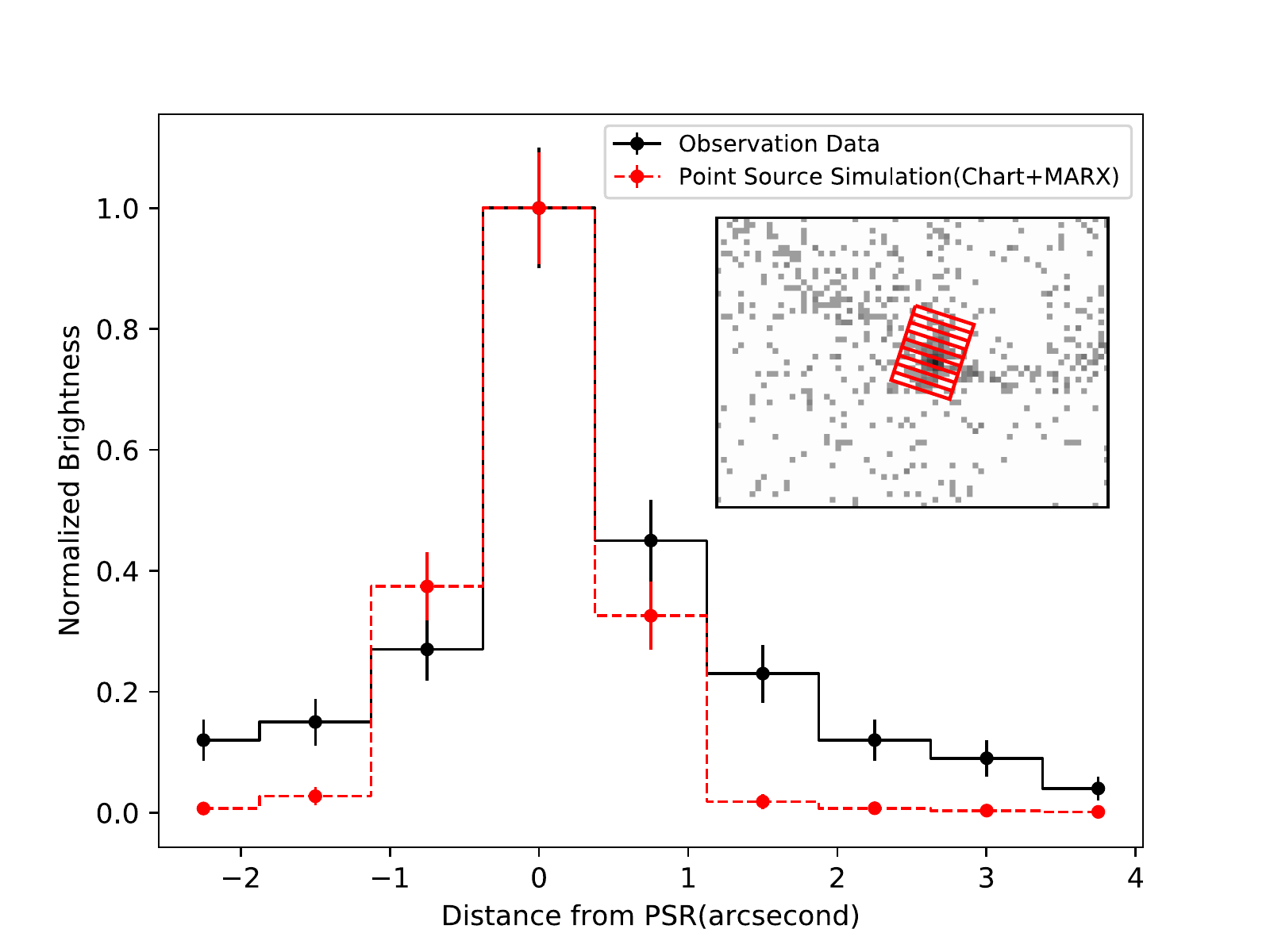}
  \caption{1a}
\end{subfigure}%
\begin{subfigure}{.33\textwidth}
  \centering
  \includegraphics[width=1.0\textwidth]{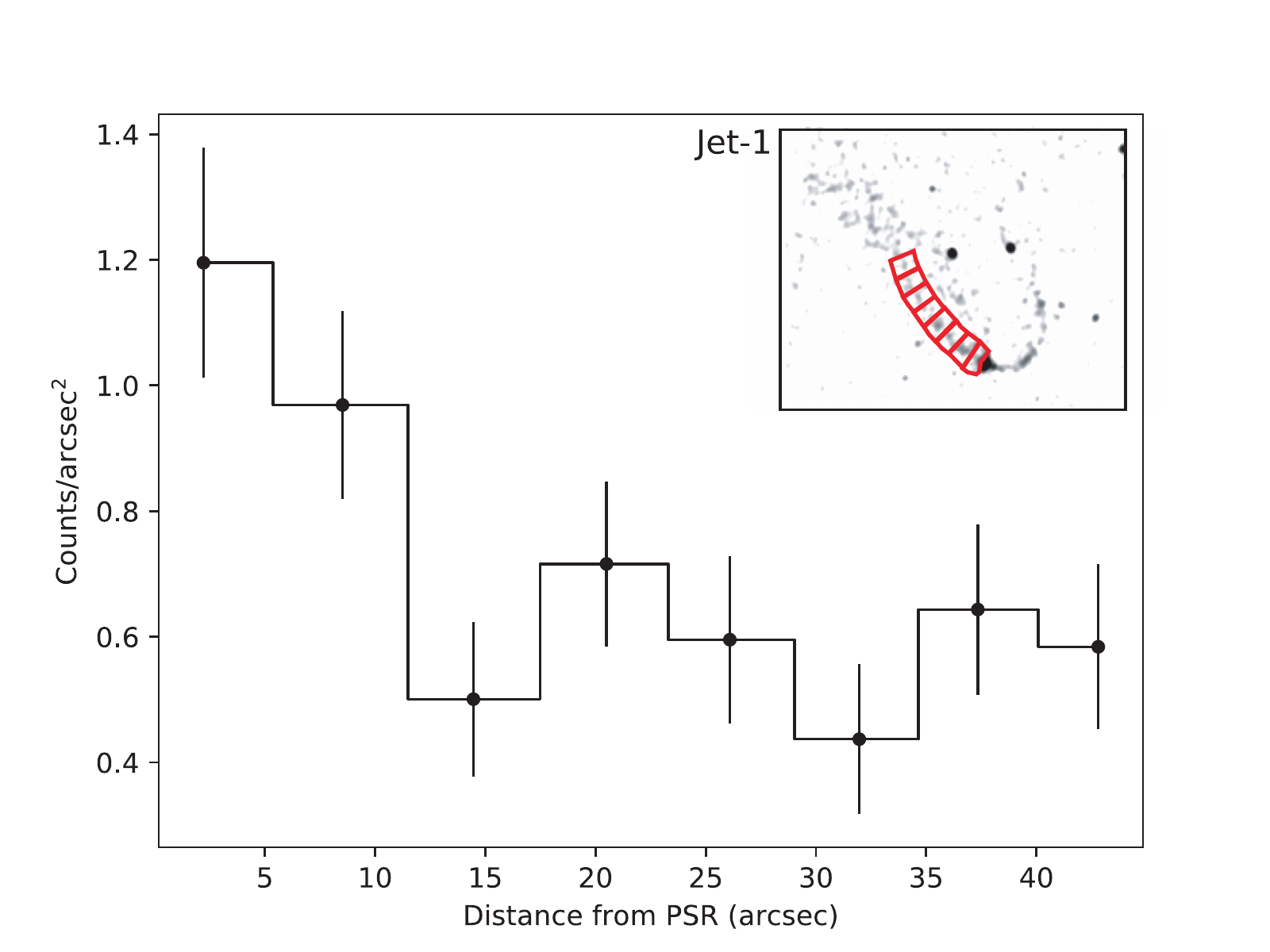}
  \caption{1b}
\end{subfigure}
\begin{subfigure}{.33\textwidth}
  \centering
  \includegraphics[width=1.0\textwidth]{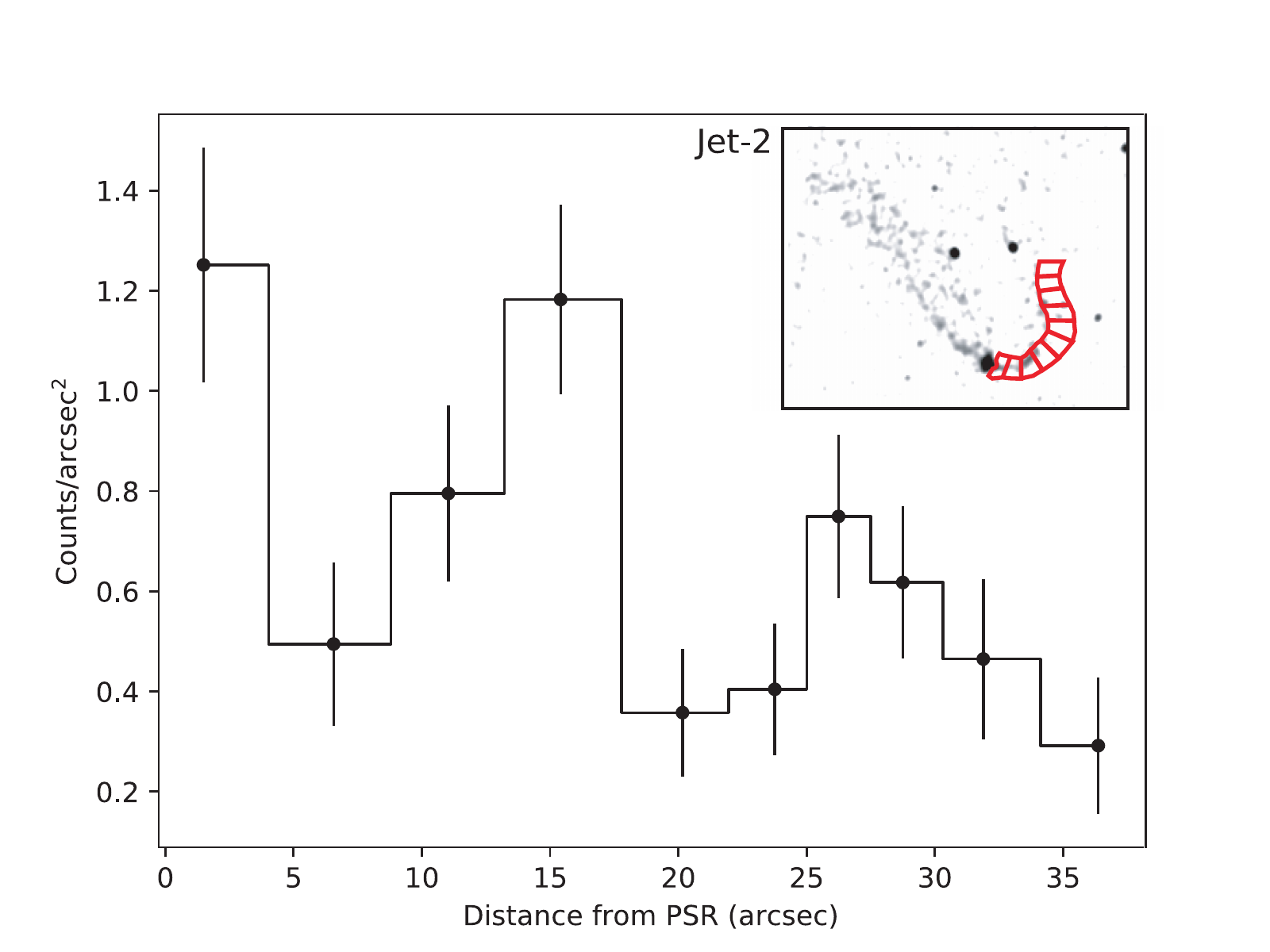}
  \caption{1b}
  \end{subfigure}
\caption{Surface brightness profiles along the compact nebula around \psr (left panel) and the eastern (jet-1, middle panel) and western (jet-2, right panel) jet-like structures. The profile of the simulated \textit{Chandra} PSF for a point-like source with spectral properties as the ones derived for \psr and normalized to its peak intensity at the location of the pulsar is also displayed in the left panel.
}
\label{flux_profiles}
\end{figure*}

{From the final flux map produced as explained in Sect.~\ref{observations}}, in addition to the PWN itself, two asymmetric jet-like structures can be clearly distinguished. The eastern jet-like feature, Jet-1, appears more diffuse and displays a much more rectilinear structure compared to the western jet (Jet-2), which displays instead an ``arc'' shape, likely related to the pressure exerted on the outflow by the surrounding medium.

In the FoV within 1 arcmin around the pulsar, there are also six point-like X-ray detected background sources, marked with a green circle in Fig.~\ref{skymap}. Five of these are present in the \textit{Chandra} Source Catalogue (S1: 2CXO J113510.4-605605, S2: 2CXO J113512.2-605559, S3: 2CXO J113514.1-605605, S4: 2CXO J113509.9-605500, and S5: 2CXO J113503.0-605521). The sixth point-like source (S6) is unassociated, and shows significant flux variability between DS-1 and DS-2 and DS-3. Further studies are needed to constrain its nature and/or any possible association at other wavelengths.

Surface brightness profiles were extracted for both Jet-1 and Jet-2. Whereas the eastern jet displays a rather smooth flux profile, the western feature displays a moderately significant enhancement about 18$\arcsec$ away from \PSR, followed by a flux decrease and recovering afterwards to a relatively smooth profile (see Fig.~\ref{flux_profiles}). These profiles further highlight the asymmetry of the two jet-like structures. A flux profile was also extracted for \psr and its close nebula using 8 rectangular slices along the south-east to north-west direction (see left panel in Fig.~\ref{flux_profiles}). The PWN displays extended emission along this axis, with a moderately broad flux profile which significantly extends beyond the \textit{Chandra}'s point spread function (PSF). {This extended emission is interpreted as the nebula powered by the central pulsar (and labeled in the following as the ``compact nebula'', CN)}.

\subsection{Results: spectra}
\label{morphology}

Spectral fits were extracted for both the PSR and Jet-1 and Jet-2 regions. Jet-1 was further subdivided into an inner and an outer regions in order to look for spectral variations, e.g. due to radiation losses. The low-counts retrieved for the CN prevents from a meaningful spectral fit to this component. {We note that some counts from the CN region may have been ascribed to the PSR region given their overlapping spatial distribution. However, given the relative low-counts from the CN region as compared to the PSR one, the effects on the final spectral results on the latter are negligible.}.

Simple power-law spectral models were employed in all cases. For the PSR, adding a black-body component worsened the fits results significantly. The best fit parameters for the PSR component are a spectral index $\Gamma_{\rm PSR} = 1.31 \pm 0.29$, a flux normalisation $N_{\rm 0, PSR} = (3.15 \pm 1.12) \times 10^{-6}$~ph~s$^{-1}$~cm$^{-2}$ and a column density of $N_{\rm H} = 0.51 \pm 0.19 \times 10^{22}$~cm$^{-2}$. For the spectral analysis of all Jet-1 and Jet-2 regions, column densities were fixed to the value obtained in the spectral fit of the PSR. Spectral indices for Jet-1 are found to be
$\Gamma_{\rm Jet1} = 1.71 \pm 0.08$ (with $\Gamma_{\rm Jet1}^{\rm inner} = 1.70 \pm 0.11$ and $\Gamma_{\rm Jet1}^{\rm outer} = 1.71 \pm 0.12$), and $\Gamma_{\rm Jet2} = 1.88 \pm 0.12$ for Jet-2. A summary of the spectral fits is reported in Table~\ref{spectral_fits}. These values are in agreement, within uncertainties, with the values reported in \citet{Marelli2012} for a similar analysis of \psr but accounting only for \textit{Chandra} observations in 2003 (DS-1 in Table~\ref{table:observations}) .

\subsection{Results: proper motion}
\label{proper_motion}

{\psr lacks an accurate estimate of its proper motion velocity. An upper limit $v_{\rm J1135, \perp} \leq 330$~km~s$^{-1}$ is provided in \citet{Kargaltsev2017}. We take advantage of the time span of the $Chandra$ observations studied here to constrain \PSR's velocity. Following the procedures outlined in Sect.~\ref{observations}, we used \texttt{wavdetect} and \texttt{wcs$\_$match} to retrieve and then match the position of \psr amongst data sets, taking DS-2 as a reference since it accounts for the longest available exposure. \psr is found at (RA, Dec) = ($11\hours 35\minutes 08.38\seconds \pm 0.02 \seconds, -60\degrees 55\arcmin 36\arcsec.49 \pm 0.14\arcsec$) in DS-1 and at (RA, Dec) = ($11\hours 35\minutes 08.41\seconds \pm 0.01\seconds, -60\degrees 55\arcmin 36\arcsec.29 \pm 0.12\arcsec $) in DS-3. Therefore, a shift of ($\Delta$RA, $\Delta$Dec)$_{\rm DS1, DS3}^{\rm J1135} = (0.035\seconds \pm 0.021\seconds, 0.20\arcsec \pm 0.19\arcsec)$ is retrieved. Considering the time span of $\sim 11.4$~yrs between DS-1 and DS-3 and applying astrometric corrections, this translates into a shift of $\mu_{\rm J1135}\approx 0.043 \pm 0.028$~arcsec~yr$^{-1}$.
In order to account for systematic effects, we apply the same procedure with the catalogued sources S3 and S5, which we take as reference sources since these are detected at a statistical significance above 5$\sigma$ in our analysis. A weighted mean angular shift for these two sources is computed between DS-1 and DS-3, weighted with the square of their positional uncertainties, which is found to be ($\Delta$RA, $\Delta$Dec)$_{\rm DS1, DS3}^{\rm S3, S5}$ = ($0.009\seconds \pm 0.023\seconds,0.463 \pm 0.186\arcsec$). This translates into a shift, after applying astrometric corrections, of $\mu_{\rm S3, S5} \approx 0.042 \pm 0.031$~arcsec~yr$^{-1}$.
Since $\mu_{\rm J1135}$ is within errors at the level of $\mu_{\rm S3, S5}$, which we take as a systematic uncertainty in our measurement, we use this latter value to derive an upper limit to \psr's velocity. Taking a distance $d_{\rm J1135}=2.9$~kpc, this translates into $v_{\rm J1135, \perp} \leq 282$~km~s$^{-1}$.
}

\begin{table*}[ht]
\centering
\caption{Spectral analysis results of jet-1, jet-2 and PSR regions. {\texttt{cstat}} statistics are used. N$_{\rm H}$ values are fixed to the value found in the spectral fit of the PSR. Errors indicate the 1$\sigma$ statistical level uncertainty.}
\begin{tabular}[t] {p{0.15\linewidth} P{0.1\linewidth} P{0.25\linewidth} P{0.2\linewidth} P{0.2\linewidth}}
\toprule
\toprule
 & $\Gamma$ 	& N$_{0}$							& N$_{\rm H}$ 					& $\chi^{2}$/d.o.f \\
 &  			& ($\times 10^{-6}$ ph~s$^{-1}$~cm$^{-2}$) 	& ($\times 10^{22}$ cm$^{-2}$) 		&  \\
\midrule
\addlinespace[0.25cm]
PSR   &  1.31 $\pm 0.29$ & 3.15 $\pm 1.12$ & 0.51 $\pm 0.19$ &  349/443 = 0.79\\
\addlinespace[0.25cm]
Jet-1 & 1.71 $\pm 0.08$ & 15.06 $\pm 1.03$ & 0.51 $\pm 0.19$ & 496/444 = 1.12  \\
\addlinespace[0.15cm]
\cmidrule [0.1pt]{1-1}
\addlinespace[0.15cm]
~~Jet-1 inner &  1.70 $\pm 0.11$ & 8.10 $\pm 0.77$ & 0.51 $\pm 0.19$ & 411/444 = 0.93  \\
\addlinespace[0.15cm]
~~Jet-1 outer &  1.71 $\pm 0.12$ & 6.93 $\pm 0.71$ & 0.51 $\pm 0.19$ & 399/444 = 0.90  \\
\addlinespace[0.15cm]
\cmidrule [0.1pt]{1-1}
\addlinespace[0.15cm]
Jet-2        &  1.88 $\pm 0.12$ & 8.02 $\pm 0.78$ & 0.51 $\pm 0.19$ & 440/444 = 0.99  \\
\addlinespace[0.25cm]
\bottomrule
\end{tabular}
\label{spectral_fits}
\end{table*}%

\section{Discussion}
\label{Discussion}

The analysis of \textit{Chandra} observations on \psr reported in Sect.~\ref{observations} reveals, in addition to the bright emission from \PSR, the presence of several extended features likely connected to it, e.g. by the shocked wind of the pulsar moving at supersonic speeds through the surrounding medium \citep{Gaensler2006, Bykov2017}. A compact nebula is found extending beyond the $Chandra$'s PSF both ahead and behind the PSR location, whereas two lateral jet-like structures displaying a highly asymmetric geometry are also clearly distinguished.

Similar large X-ray extended features have been observed in a few other PWNe (\citealp{Kargaltsev2017}). The morphology of \psr is reminiscent of that observed in the runaway pulsars PSR J1509--5850 \citep{Klingler2016} and Geminga \citep{Posselt2017}, both displaying an axial tail and two lateral outflows. Assuming a distance to \psr of $d_{\rm J1135}=2.9$~kpc \citep{Marelli2012}, {similar physical extensions of these structures are also recovered: 0.05 pc and 0.13--0.18 pc in PSR J1509--5850 (or 2.7$\arcsec$ and 7--10$\arcsec$ at a distance $d_{\rm J1509}=2.9$~kpc) ; 0.07 pc and $\sim$ 0.3 pc in Geminga (or 1.2$\arcmin$ and 5$\arcmin$ assuming $d_{\rm Geminga}=0.25$~kpc); and 0.04 pc and $\sim$ 0.5--0.7 pc in \PSR (or 2.8$\arcsec$ and 0.5--1$\arcmin$ for $d_{\rm J1135}=2.9$~kpc), for the tail and jet-like features, respectively}.




The origin of the extended emission in front of \PSR, and the two lateral outflows is uncertain. We tentatively interpret the extended emission ahead of \psr as the region encompassed by the bow shock in a sPWN. Its extension, of about 2 arcsec or $\sim 8.7 \times 10^{16}$~cm at a distance $d_{\rm J1135} = 2.9$~kpc, should be of the order of the stand-off radius $R_{\rm S}$ and can be used to infer the proper motion velocity of the system:
$v_{\rm psr} \approx ( \dot{E}_{\rm PW}\, ( 1/4 \, \pi \, c \, \rho_{\rm ISM} R_{\rm S}^{2})^{1/2}$ $\sim 210$~km~s$^{-1}$,
taking $\dot{E}_{\rm PW} = 2.1 \times 10^{36}$~erg~s$^{-1}$ and an ISM particle density of $n_{\rm ISM} = 1$~cm$^{-3}$.
This estimate is in accordance with {the value obtained in Sect.~\ref{proper_motion}, $v_{\rm psr}\leq 282$~km~s$^{-1}$}. We have assumed here that the entire spin-down power of \psr is carried away by the pulsar wind ($\xi = 1$), which is in turn taken to be isotropic for simplicity ($f_{\rm iso} = 1$). The value of $v_{\rm psr}$ depends on these parameters as $v_{\rm psr} \propto (\xi \, f_{\rm iso})^{1/2}$ and can therefore change within factors of $\sim$ a few.
The physical origin of the CN emission {around} the PSR would correspond in this scenario {to the pulsar equatorial outflow shocked by the ISM ram pressure either ahead or trailing he pulsar proper motion}. As for the lateral outflows, they would correspond to bipolar jets, again bent by the external pressure of the medium. The asymmetry observed between the eastern/western outflows could be due to their different direction of propagation with respect to the pulsar proper motion, to a different relative kinetic power injected into each outflow, or to different properties of the ISM through which they propagate. Similar scenarios have been proposed for the Geminga PWN, \citep{Pavlov2006}. This interpretation is also in accordance with recent numerical simulations of extended outflows originated in fast-moving PWNe \citep{Barkov2019b}.
The behaviour obtained for \PSR's outflows resembles instead the ones found in PSR\,J1509--5850, displaying an index $\sim 1.8$ and 1.9. Both Geminga and PSR\,J1509--5850 display also an axial tail trailing the pulsar, with spectral indices $\Gamma \gtrsim 2.0$ and 1.4, respectively. Unfortunately, the low statistics obtained in this report for the CN region around \psr prevents any further comparison of the tail spectral properties.

In this scenario, the observed X-ray emission would be produced by high-energy electrons embedded in the jets' magnetic field $B_{\rm jet}$. The absence of any significant spectral break due to synchrotron cooling along the jets (particularly for the inner and outer regions of Jet-1) implies that $\tau_{\rm sync} \approx 100 \, (E_{\rm ph}/{\rm 1\,keV})^{1/2}\,(B_{\rm jet}/50\,{\rm \mu G})^{3/2} \geq \tau_{\rm dyn} \sim l_{\rm jet}/v_{\rm jet}$, where $l_{\rm jet}$ and $v_{\rm jet}$ are the jet length and flow velocity, respectively. A lower limit on the bulk velocity of the jet outflows can thus be placed,  $v_{\rm jet} \gtrsim 8000 $\,km~s$^{-1}$. The fact that electrons are confined within the jet, on the other hand, implies that their gyroradius $r_{\rm gyro} = \gamma_{\rm e}\,m_{\rm e}c^{2} / eB_{\rm jet}$ cannot exceed the jet radius, $R_{\rm jet}$. For a given magnetic field $B_{\rm jet}$, this condition can be used to derive a maximum Lorentz factor of the emitting electrons along the jet, $\gamma_{\rm e}^{\rm max}$. A minimum value, $\gamma_{\rm e,min}^{\rm X}$, is on the contrary needed for synchrotron emission to reach the keV band for the same value of $B_{\rm jet}$. Putting these limits together, one obtains $6.3 \times 10^{7} (B/50\,{\rm \mu G})^{-1/2} \leq \gamma_{\rm e}^{\rm X}\leq 3.7 \times 10^9 (B/50\,{\rm \mu G})$.



An alternative interpretation for the CN in \psr and the two lateral outflows may also be envisaged. The latter could represent the projection of a limb-brightened shell formed in the region of the contact discontinuity separating the pulsar shocked wind with the shocked ISM material. The CN around the PSR could correspond in this scenario to a pulsar jet launched along the pulsar’s spin axis. Given the moderately high speed of \PSR, such a forward jet would propagate ahead rather shortly, braked and eventually deflected by the strong pressure exerted by the ISM, whereas it should instead be able to propagate up to longer distances behind the PSR. In this regard, it is worth noting that the extended emission observed from the long, eastern outflow Jet-1 is relatively wider than the western jet-like outflow, and quite non-homogeneous (see Fig.~\ref{skymap}). In particular, a few arcsec away from the position of the pulsar, the Jet-1 seems to be divided in two broad quasi-rectilinear structures, separated by a region displaying a comparatively lower surface brightness. At larger distances, these structures smoothly converge again into a diffuse, wider structure. While the Jet-1 inner and outer regions may still not represent a proper jet but the limb-brightened shell of the PWN, the inner outflow could be attributed to emission produced by a backwards-propagating jet, slightly deflected/bent by the external medium pressure, which at large distances approaches the PWN shell eastern limb. Note however that a similar structure may also be produced by a ram-pressure confined PWN tail. In that case, however, a bending may not be expected, as the tail should propagate along the direction of motion, confined by an almost symmetrical lateral pressure. Deeper observations of the PWN around \PSR are in any case needed to conclusively assess whether this eastern outflow is indeed composed by several extended sub-structures (see, e.g. \citealp{Pavan2016}).

The spectral and morphological properties of the extended structures discussed above cannot exclude alternative scenarios for the production of the jet-like structures {observed in \PSR }. Assuming that the lateral outflows correspond to true pulsar jets, a significant bending may be expected and is indeed clearly observed {in our morphological analysis, particularly for the western outflow}. This is contrary to more extreme jet-like features seen in other sPWN (e.g. in the Guitar Nebula \citep{Hui2012} and the Lighthouse Nebula \citep{Pavan2014}), in which the length and orientation of these features favours a scenario based on the diffusion of high-energy escaping electrons {(or ``kinetic jets'' \citealp{Barkov2019a})} from the sPWN. {On spectral grounds, our analysis could not retrieve any softening along the jet-like features which could constrain the energy distribution of the underlying emitting particle population (see e.g. \citep{Bandiera2008})}. Further observations of \PSR, both in X-rays or at lower wavelengths (e.g. in the radio band) may be able to resolve such a spetral fingerprint.



\vspace{2cm}

\begin{acknowledgements}
PB and XZ acknowledge the financial support by the Spanish Ministerio de Econom\'{i}a, Industria y Competitividad (MINEICO/FEDER, UE) under grant AYA2016-76012-C3-1-P, from the State Agency for Research of the Spanish Ministry of Science and Innovation under grant PID2019-105510GB-C31 and through the “Unit of Excellence Mar\'{i}a de Maeztu 2020-2023” award to the Institute of Cosmos Sciences (CEX2019-000918-M), and by the Catalan DEC grant 2017 SGR 643.
\end{acknowledgements}

\bibliographystyle{aa}
\bibliography{bibliography}

\end{document}